\documentclass{format}
\usepackage{amsmath,amsxtra,amssymb,latexsym, amscd,color}
\usepackage[mathscr]{eucal}
\usepackage[super,compress]{cite}
\usepackage[T1]{fontenc}
\usepackage[utf8]{inputenc}
\usepackage{lmodern}
\usepackage{hyperref}

\def\pA{proton-nucleus\ }

\def\pG{$(p,\gamma)$\ }
\def\C3p{$p+^{13}$C\ }
\def\C2p{$p+^{13}$C\ }
\def\pC23{$p+^{12,13}$C\ }

\begin{document}

\markboth{N. H. Phuc, N. T. T. Phuc \& D. C. Cuong}{Study of nonlocality effects in direct capture reactions with Lagrange-mesh $R$-matrix method}

\catchline{}{}{}{}{}

\title{Study of nonlocality effects in direct capture reactions with Lagrange-mesh $R$-matrix method}
\author{Nguyen Hoang Phuc}

\address{Institute for Nuclear 
 Science and Technology, VINATOM\\
179 Hoang Quoc Viet, Cau Giay, Hanoi, Viet Nam\\
hoangphuc@vinatom.gov.vn }



\author{Nguyen Tri Toan Phuc\footnote{Corresponding author}}

\address{Department of Nuclear Physics, Faculty of Physics and Engineering Physics, \\ University of Science, \\
227 Nguyen Van Cu, District 5, Ho Chi Minh City, Viet Nam\\
nttphuc@hcmus.edu.vn }
\address{Vietnam National University, Ho Chi Minh City, Viet Nam}

\author{Do Cong Cuong}

\address{Institute for Nuclear 
 Science and Technology, VINATOM\\
179 Hoang Quoc Viet, Cau Giay, Hanoi, Viet Nam\\
cuong1981us3@gmail.com}
\maketitle

\begin{history}
\received{Day Month Year}
\revised{Day Month Year}
\end{history}

\begin{abstract}
We apply the Lagrange-mesh $R$-matrix method to calculate the $S$-factor for the $^{13}$C$(p,\gamma)^{14}$N and $^{16}$O$(p,\gamma)^{17}$F direct radiative capture reactions. By comparing the astrophysical $S$-factors calculated with nonlocal and local potentials, we investigate the nonlocality effects coming from the nuclear potentials in the direct capture reactions. Our calculations are in good agreement with the experimental data and indicate a nonnegligible difference in the results of local and nonlocal potentials. The use of small diffuseness narrow potentials also provides a remarkably good fit in the case with multiple broad resonances. Our findings suggest that the nonlocal potential improves the calculated results although the difference between the local and nonlocal potentials is smaller than uncertainties from other sources. We propose the nonlocality potential should be used in the potential model calculation of future astrophysics rates evaluation.      
\end{abstract}


\ccode{PACS numbers: 25.40.Lw, 25.40.Ny, 26.20.Cd, 21.10.Jx }

\section{Introduction}

Nuclear radiative capture reactions play a critical role in various astrophysical processes. An important goal in nuclear astrophysics is to determine rates for radiative capture reactions that take part in nucleosynthesis processes. For those processes involving light nuclei at low energies, the direct capture mechanism dominates over the compound nucleus one. These direct radiative capture reactions can be found in many processes such as the big bang nucleosynthesis, stellar evolution, and element synthesis at many astrophysical sites \cite{Rol88,Ber16}. Therefore, the reaction models used to describe the direct radiative capture cross section and their validity are extensively and actively researched subjects of nuclear astrophysics. 

There are several different approaches to describe the direct capture at sub-Coulomb barrier energies for light nuclei of astrophysical interest \cite{Des20}. At the two extremes are the phenomenological $R$-matrix method \cite{Des10,Azu10,Duc21} and the microscopic models \cite{Des20}. While the former approach is used as an effective way to parameterize the reaction in terms of resonance property, the latter one, which includes the \emph{ab initio} \cite{Bar13} and microscopic cluster models \cite{Wil77,Lan94}, aims to describe the reaction starting from the nucleon degree of freedom. Both approaches have been extensively studied and used in the past. The third approach is the potential model \cite{Chr61,Hua10,NACRE2,Dub17,Des20}, which lies in between the two extremes and is also widely used by the nuclear astrophysics community. The advantage of the potential model is its flexibility while still keeping a microscopic connection through the use of nuclear potential. In this work, we use the potential model to describe the radiative capture reactions.

The nuclear potentials, which describe the scattering and binding of the projectile nucleon with the target, are in principle nonlocal in coordinate space. It is well known that the main source of the nonlocality in these potentials originates from the antisymmetrization due to the Pauli principle and channel couplings \cite{Fra08}. Nonlocal potentials have been investigated in nuclear physics since the seminal works \cite{Bet56,Fra57,Per62} in the 1950s and 60s. In recent years, the topic of nonlocal potential has been revived and its effect has been studied in various types of reactions such as elastic scattering \cite{Jag18,Mah14}, transfer \cite{Del09,Tim13,Tit16,Li18}, breakup \cite{Gal98}, proton \cite{Ter16} and $\alpha$ \cite{Per19} decays, and fusion \cite{Gal94,Bai21}. 

The nonlocality effect in the potential model of direct capture reaction has only been studied recently by Tian, Pang, and Ma \cite{Tia18}. In that work, the Schr\"{o}dinger equations with the Perey-Buck-type nonlocal potentials for both bound and scattering state were solved by applying the Numerov method iteratively. The $^7$Li($n,\gamma)^8$Li, $^{12}$C$(p,\gamma)^{13}$N, and $^{48}$Ca($n,\gamma )^{49}$Ca cases considered in Ref.~\citen{Tia18} only contain experimental data for the total capture cross sections although the authors of that work did calculate the theoretical transition to the ground and excited states. They reported a considerable nonlocality effect of 20\%-25\% for the nonresonant $^7$Li($n,\gamma)^8$Li, $^{48}$Ca($n,\gamma )^{49}$Ca reactions, and the $^{12}$C$(p,\gamma)^{13}$N reaction, which contains one broad resonance in the considered energy range. Since many essential direct capture reactions exhibit multiple overlapping resonances or contain large contributions from the transition to excited states, an evaluation of the nonlocality effect in these cases is of high interest.       

In this study, we focus on the $^{13}$C$(p,\gamma)^{14}$N and $^{16}$O$(p,\gamma)^{17}$F reactions. The $^{13}$C$(p,\gamma)^{14}$N reaction at below Coulomb barrier energies is a key process in the CNO cycle of hydrogen burning stage \cite{Rol88,Ber16}. This reaction also depletes the $^{13}$C nucleus, which is the seed for the neutron-producing reaction $^{13}$C$(\alpha,n)^{16}$O in the asymptotic giant branch (AGB) stars \cite{Lug03}. The capture cross section of $^{13}$C$(p,\gamma)^{14}$N leading to ground state in the low energy region displays two pronounced resonances at $E=0.51$ MeV \cite{Gen10,Kin94} and 1.23 MeV \cite{Zep95}, respectively (see Refs.~\citen{Cha15,Cha19} for examples of phenomenological fits). A simultaneous description of these two resonances in the $^{13}$C$(p,\gamma)^{14}$N reaction within a purely potential model has not been achieved so far. 

On the other hand, the $^{16}$O$(p,\gamma)^{17}$F at the astrophysical-interesting energies is an excellent example for a nonresonant extranuclear capture reaction, where the proton is directly captured at distance much larger than $^{16}$O radius without forming any resonance \cite{Rol73,Bay98}. The $^{16}$O$(p,\gamma)^{17}$F is a link to the higher cycle II branches of the CNO bi-cycle \cite{Rol88,Ber16}. This reaction also takes part in the AGB phase, where its reaction rate sensitively affects the $^{17}$O/$^{16}$O isotopic ratio. The experimental data for the $^{16}$O$(p,\gamma)^{17}$F reaction are precisely measured for the transition to the ground and first excited states \cite{Rol73,Mor97}.   

In the present work, we applied the ``calculable'' $R$-matrix method on a Lagrange mesh \cite{Des10,Des16}. In the $R$-matrix method, the configuration space is divided into the internal and external regions, determined by the channel radius $a$. The wave functions in the internal region are expanded over a basis. In the external region, the scattering wave functions are in their asymptotic form, and the matching with internal wave functions at the channel radius provides the scattering and $R$ matrices. The Lagrange mesh corresponds to specific bases, combined with the Gauss quadrature greatly simplifies the calculation \cite{Bay14}. This technique was successfully applied to solve the bound and scattering problems with the nonlocal potential in Refs.\citen{Hes02,Loa18}. The strength of the Lagrange-mesh $R$-matrix method lies in its ability to efficiently calculate the matrix element, straightforwardly handle the closed channels and resonances, as well as treat the nonlocal potential on the same footing with the local one within a consistent coupled-channels framework \cite{Des10}. 

Our paper is organized as follows. In Sec.~\ref{sec2}, the potential model for direct radiative capture reaction is briefly presented. In Sec.~\ref{sec3}, we describe the Lagrange mesh $R$-matrix method for the calculations of bound and scattering wave functions. In Sec.~\ref{sec4}, the formalism is applied to the $^{16}$O$(p,\gamma)^{17}$F and $^{13}$C$(p,\gamma)^{14}$N reactions using the local and nonlocal potentials. Finally, the results are summarized in Sec.~\ref{sec5}.

\section{Potential model for the radiative capture reaction}
\label{sec2}
Here we briefly introduce the general formalism of the potential model for direct radiative capture reactions with electric transition, more details can be found in Refs.~\citen{Des10,Rol73,Anh21}. The total cross section of the $A(p,\gamma)B$ reaction at the energy in the center-of-mass frame $E\equiv E_\text{c.m.}$ is obtained by summing the partial cross sections over all initial (scattering) states $|J_i M_i\rangle$ , final (bound) states $|J_f M_f\rangle$, and electric multipoles $\lambda$
\begin{equation}
\sigma(E)=\sum_{\lambda J_i J_f }\sigma_{\lambda,J_i\to J_f}(E), \label{eq1}
\end{equation}
with
\begin{align}\label{eq2}
\sigma_{\lambda,J_i \to J_f}(E)=\frac{4\pi(\lambda+1)(2\lambda+1)}{\lambda
\left[(2\lambda+1)!!\right]^2}\frac{\mu c^2}{(\hbar c)^2}\frac{k_\gamma^{2\lambda+1}}
{k^3}C_\lambda^2\frac{(2J_i+1)(2J_f+1)}{(2S+1)}S_{\text{F}} \nonumber\\
 \times\sum_{\ell_i,j_i}\left| \hat{j_i}\hat{j_f}\hat{\ell_i}
 \left\{\begin{matrix}
   j_i&J_i& S  \\
   J_f& j_f&\lambda
  \end{matrix} \right\}
	\left\{\begin{matrix}
   \ell_i&j_i& \frac{1}{2}  \\
   j_f& \ell_f&\lambda
 \end{matrix}\right\}\langle \ell_i0,\lambda0\vert \ell_f0 \rangle I(k)
 \right|^2 ,
 \end{align}
where $\mu=m_Am_p/(m_A+m_p)$ is the reduced mass of the system, $m_A$ and $m_p$ stand for the masses of the target nucleus $A$ and the proton. $C_\lambda=e\left[m_A^\lambda+Z_A(-m_p)^\lambda\right]/(m_A+m_p)^\lambda$ is the effective charge where $Z_A$ is the charge number of $A$ nucleus. $S$ is the spin of the target and $S_{\text{F}}$ is the spectroscopic factor of the bound state. We denote $\hat{j}=\sqrt{2j+1}$ where $j_f=|\ell_f \pm 1/2|$ and $j_i = |\ell_i \pm 1/2|$ are the total angular momenta of proton where $\ell_f$ and $\ell_i$ are the relative orbital angular momenta of the initial and final state, respectively. $k$ and $k_\gamma$ are the wave numbers of the proton energy $E$ and the emitted photon energy $E_\gamma$, respectively. 

In equation (\ref{eq2}), $I(k)$ is the radial overlap integral of the scattering wave function $\chi_{\ell_ij_i}$ and the normalized bound wave function $\phi_{n_f \ell_fj_f}$ ($n_f$ is the number of nodes) determined as
\begin{equation} \label{eq3}
	I(k)=\int_0^\infty\phi_{n_f\ell_fj_f}(r)\chi_{\ell_i j_i}(k,r)r^\lambda dr. 
\end{equation}
The scattering wave function $\chi_{\ell_i j_i}$ and the bound wave function $\phi_{n_f \ell_f j_f}$ are obtained by using the Lagrange-mesh $R$-matrix method described in the next section. The scattering wave function $\chi_{\ell_i j_i}$ in our calculations is normalized to have the asymptotic forms
\begin{equation}
	\chi_{\ell_i j_i}(k,r)=[F_{\ell_i}(kr) \cos\delta_{\ell_i j_i} 
	+G_{\ell_i}(kr) \sin\delta_{\ell_i j_i}] \ \text{for} \ r\to\infty, \label{eq3b}
\end{equation}
where $\delta_{\ell_i j_i}$ is the nuclear phase shifts while $F_{\ell_i}$ and $G_{\ell_i}$ are the regular and irregular Coulomb functions, respectively \cite{Abra65}.

For the $A(p,\gamma)B$ reaction at energies below the Coulomb barrier, the cross section $\sigma(E)$ decreases too rapidly when the energy $E$ decreases, therefore it is more convenient to use the astrophysical $S$-factor defined as 
\begin{equation} 
	S(E) = E\exp(2\pi\eta)\sigma(E), \label{eq4}
\end{equation}
where $\eta=Z_A e^2/(\hbar v)$ is the Sommerfeld parameter with the proton (relative) velocity $v$. 

For the local potential, the bound and scattering wave functions are determined from the solution of the single-channel radial Schr\"odinger equation 
\begin{eqnarray}\label{eq5}
	&-&\frac{\hbar^2}{2\mu}\left[\frac{d^2}{dr^2}-\frac{\ell(\ell+1)}{r^2}\right]\psi_{\ell j}(k,r)+ \left[V_L(r)+A_{\ell j}V_{\rm SO}(r)+V_{\rm C}(r)
	\right]\psi_{\ell j}(k,r)\nonumber\\
	&=&E\psi_{\ell j}(k,r),
\end{eqnarray}
where we denote the solution $\psi_{\ell j}(r)$ for the scattering state ($E>0$) as $\chi_{\ell_ij_i}(r)$ and for the bound state ($E<0$) as $\phi_{n_f \ell_f j_f}(r)$. The coefficient $A_{\ell j}=[j(j+1)-\ell(\ell+1)-3/4]$ is originated from the spin-orbit coupling between the proton and the relative orbital angular momentum. $V_L(r)$ and $V_{\rm SO}(r)$ are the central and spin-orbit parts of the local potentials, respectively. $V_{\rm C}(r)$ is the Coulomb potential in the commonly used form
\begin{eqnarray}
	V_{\rm C}(r)=\left\{\begin{array}{lcr}
		\displaystyle\frac{Z_Ae^2}{r},& & r > R_{\rm C}\\
		\displaystyle\frac{Z_Ae^2}{2R_{\rm C}}\left(3-\frac{r^2}{R_{\rm C}^2}\right),& & 
		r\leqslant R_{\rm C}, \label{eq6}
	\end{array}\right. 
\end{eqnarray}
which assumes a uniform charge distribution inside the radius $R_{\rm C}= r_{\rm C}A^{1/3} = 1.25~A^{1/3}$ (fm) with $A$ being the mass number of the target.

In the potential model, the phenomenological Woods-Saxon (WS) and its derivative forms are commonly used for the central and spin-orbit potentials, respectively
\begin{align} 
	V_{\rm L}(r) &= -V_{\rm L}f_{\rm c}(r), \label{eq7}\\
	V_{\rm{SO}}(r) &= V_{\rm SO} \left(\dfrac{\hbar}{m_\pi c}\right)^2 
	\dfrac{1}{r} \dfrac{d}{dr} f_{\rm SO}(r),\label{eq8} \\
	{\rm where}\ f_x(r) &=\left[1+\exp\left(\dfrac{r-R_x}{a_x}\right)\right]^{-1}, 
	\ x={\rm c},{\rm SO}. \label{eq9}
\end{align}
Here $R_x$ and $a_x$ are the radius and diffuseness of the WS potential, 
respectively, while $[\hbar/(m_\pi c)]^2\approx 2$ fm$^2$.

For the nonlocal potential calculation, we still use the local form for the spin-orbit and Coulomb terms and only consider the central term in the Perey-Buck form of nonlocality \cite{Per62}. The Schr\"{o}dinger equation reads
\begin{eqnarray}\label{eq10}
	-\frac{\hbar^2}{2\mu}\left[\frac{d^2}{dr^2}-\frac{\ell(\ell+1)}{r^2}\right]\psi_{\ell j}(k,r)+ \left[V_{\rm C}(r)+
	A_{\ell j}V_{\rm SO}(r)\right]\psi_{\ell j}(k,r)\nonumber\\
	+\int v_\ell(r,r')\psi_{\ell j}(k,r')dr'=E\psi_{\ell j}(k,r),
\end{eqnarray}
where
\begin{equation}\label{eq11}
	v_\ell(r,r')=U\left(\frac{r+r'}{2}\right)\frac{1}{\pi^{\frac{1}{2}}\beta}\exp\left[-\frac{(r^2+r'^2)}{\beta^2}\right]2i^\ell z j_\ell(-iz),
\end{equation}
with the WS form $U(x)=V_{\rm NL}f_{\rm c}(x)$ and $z=2rr'/\beta^2$, $\beta$ is the range of the nonlocality and $j_\ell(x)$ is the spherical Bessel function. 

In general, the \pG radiative capture cross section (\ref{eq1}) contains both the resonant and nonresonant processes. While the resonant scattering wave 
function in the potential model is generated by the \pA potential accurately fine-tuned to reproduce the location of the resonance peak, the nonresonant contribution to the \pG cross section varies smoothly with the incident energy $E$ and is closely related to the asymptotic expression of the final (bound) state. It is known that the nonresonant \pG cross section is proportional to $S_{\text{F}} b^2_{n_f\ell_fj_f}$ \cite{Muk97,Muk01}. Here $b_{n_f\ell_fj_f}$ is the amplitude of the asymptotic tail of the bound state wave function and is called the single-particle Asymptotic Normalization Coefficient (ANC), whose calculation is described in the next section. Therefore, it is more convenient to characterize the nonresonant capture cross section in terms of the final state ANC determined \cite{Muk97,Muk01} as 
\begin{equation}
	C_{F} = S_{\text{F}}^{1/2}b_{n_f\ell_fj_f}. \label{eq15}
\end{equation}
In the valence nucleon + core prescription used in the present work for the daughter 
nucleus, the spectroscopic factor $S_{\text{F}}$ is chosen to match the magnitude of the experimental cross section.

\section{The Lagrange-mesh $R$-matrix method}
\label{sec3}

The calculable $R$-matrix method \cite{Des10,Des16} is a technique to solve the Schr\"{o}dinger equation, which is traditionally done by finite-difference methods such as the Numerov algorithm. Although sharing the same fundamental \cite{Lan58}, the approach of the calculable $R$-matrix method is different from that of the phenomenological $R$-matrix method \cite{Azu10,Duc21}, which is a way to effectively parameterize various types of cross sections. The fundamental principle of solving the Schr\"odinger Eq.~(\ref{eq5}) and (\ref{eq10}) using the $R$-matrix method is that the configuration space is divided at the channel radius $a$ into an internal region and an external region. The channel radius is chosen large enough to assume that the nuclear potential vanishes in the external region. In the internal region, the wave function can be expanded over some basis involving $N$ linearly independent function $\varphi_n$ as
\begin{equation}\label{eq16}
	\psi_{\ell j}^{\rm int}(r)=\sum_{n=1}^{N}c_n\varphi_n(r),
\end{equation}
where ${\varphi_n}$ is a set of basis functions. Depending on the asymptotic form of external wave function, matching it with the internal one at the channel radius will provide the solution for bound and scattering problems. In this section we present the formula to solve the Schr\"odinger equation for bound and scattering states with the case of nonlocal potential. The case with local potential can be naturally derived in the same manner.

\subsection{The scattering state}

In framework of the calculable $R$-matrix method for the scattering problem, the internal and external radial functions are connected at the channel radius $r=a$ through continuity conditions. This leads to the definition of $R$-matrix at a given energy $E$ as
\begin{equation}\label{eq17}
	\psi_{\ell j}(a)=\mathscr{R}_{\ell j}(E)[a\psi^{'}_{\ell j}(a)-B\psi_{\ell j}(a)],
\end{equation}
where $B$ is the dimensionless boundary parameter. The $R$ matrix has dimension of 1 in the single-channel case considered in this work and it is just a function of energy $E$. 

The Hamiltonians in Eqs.~\eqref{eq5} and \eqref{eq10} are not Hermitian over the internal region $(0, a)$. This situation can be remedied by introducing the Bloch surface operator \cite{Blo57,Rob69}
\begin{equation}\label{eq18}
	\mathscr{L}(B)=\frac{\hbar^2}{2\mu}\delta(r-a)\left(\frac{d}{dr}-\frac{B}{r}\right),
\end{equation}
so that the combination of the Hamiltonian and Bloch operator is Hermitian over $(0, a)$ when $B$ is real. It is well known that for the scattering problem, the results obtained using the $R$-matrix method are independent of $B$. Therefore, we have fixed $B = 0$ for the scattering state calculation and the Schr\"odinger equation in the internal region is approximated by the inhomogeneous Bloch-Schr\"odinger equation. The radial equation Eq.~(\ref{eq10}) with the nonlocal potential can now be rewritten as
\begin{eqnarray}
	\left\{-\frac{\hbar^2}{2\mu}\left[\frac{d^2}{dr^2}-
	\frac{\ell(\ell+1)}{r^2}\right]+V_C(r)+A_{\ell j}
	V_{\rm SO}(r)-E+\mathscr{L}(0)\right\}\sum_{i=1}^Nc_n\varphi_n(r)\nonumber\\
	+\int v_\ell(r,r')\sum_{i=1}^Nc_n\varphi_n(r')dr'=\mathscr{L}(0)\psi_{\ell j}^{ext}(r).
	\label{eq19}
\end{eqnarray}
The Bloch operator ensures the continuity of the derivative of the wave function. Projecting both sides of Eq.~(\ref{eq19}) on $\varphi_i(r)$ and integrating over $r$ variable, we obtain the equation as
\begin{equation}\label{eq20}
	\sum_{n=1}^NC_{in}(E,0)c_n=\frac{\hbar^2}{2\mu}\varphi_i(a)\frac{d\psi_{\ell j}^{ext}(r)}{dr}\big{\vert}_{r=a},
\end{equation}
with $C_{in}(E,0)$ denoting the matrix elements on the left-hand side,
\begin{eqnarray}\label{eq21}
	C_{in}(E,0)= && \int\varphi_i(r)\left\{-\frac{\hbar^2}{2\mu}\left[\frac{d^2}{dr^2}-
	\frac{\ell(\ell+1)}{r^2}\right]+V_C(r)+A_{\ell j}
	V_{\rm SO}(r)-E+\mathscr{L}(0)\right\}\varphi_n(r)dr \nonumber\\
	&+& \int \varphi_i(r)v_\ell(r,r')\varphi_n(r')drdr'.
\end{eqnarray}
In the Lagrange-mesh method used in this work, the basis functions $\varphi_n(r)$ are chosen as modified Lagrange functions, which are very convenient for the calculation of $C_{in}(E,B)$ (see the explicit expressions in section 4.2 of Ref.~\citen{Des10}). With this method, the integral of a function can be calculated by the sum of the function values at the mesh points $\{r_i\}$, which relates to the solutions of the Legendre polynomial multiplying with the weight $\lambda_i$ of the Gauss-Legendre quadrature in the interval $[0,a]$. The basic functions satisfy the Lagrange condition \cite{Des10},
\begin{equation}
	\varphi_n(r_i)=(\lambda_i)^{-1/2}\delta_{in}\label{eq27}.
\end{equation} 
Because of the property (\ref{eq27}), the calculation is strongly simplified and this leads to the important results of the matrix element of the local potential
\begin{equation}\label{eq28}
	\int_0^a \varphi_i(r) V(r)\varphi_n(r)dr= \sum_{k=1}^N \lambda_k \varphi_i(r_k)V(r_k)\varphi_n(r_k)= V(r_i)\delta_{in},
\end{equation}
and that of nonlocal potential
\begin{equation}\label{eq29}
	\int \varphi_i(r) v_\ell(r,r') \varphi_n(r') dr'dr = \sqrt{\lambda_i\lambda_n} v_\ell(r_i,r_n),
\end{equation}
here $V(r)$ stands for all local terms of Eq.~({\ref{eq19}).

Solving system of linear equations \eqref{eq20} gives the coefficients $c_n$. Inserting them into (\ref{eq16}) at $r=a$ and using (\ref{eq17}), one obtains the $R$ matrix
\begin{equation} \label{eq22}
	\mathscr{R}_{\ell j}(E)=\frac{\hbar^2}{2\mu a}\sum_{i,n=1}^N \varphi_i(a)(\mathbf{C}^{-1})_{in}\varphi_n(a),
\end{equation}
where $\mathbf{C}$ is the symmetric matrix whose elements defined in Eq.~\eqref{eq21}.

The external wave function $\psi_{\ell j}^{\rm ext}$ is determined as the asymptotic form of a scattering wave function as described in Eq.~\eqref{eq3b}
\begin{equation}
	\chi^{\rm ext}_{\ell j}(k,r)\equiv\psi^{\rm ext}_{\ell j}(r) = [F_{\ell}(kr) \cos\delta_{\ell j} 
	+G_{\ell}(kr) \sin\delta_{\ell j}]. \label{eq12}
\end{equation}
The external wave function contains the scattering matrix element through the nuclear phase shift as $S_{\ell j}=e^{2i\delta_{\ell j}}$. Through the continuity conditions of wave functions, the scattering matrix $S_{\ell j}$, which is sometimes referred to as the collision matrix, can be calculated from $\mathscr{R}_{\ell j}(E)$ as
\begin{equation}\label{eq24}
	S_{\ell j}=e^{2i\phi_\ell}\frac{1-L_\ell^*\mathscr{R}_{\ell j}(E)}{1-L_\ell\mathscr{R}_{\ell j}(E)},
\end{equation}
where
\begin{equation}\label{eq25}
	L_\ell=\frac{ka}{O_\ell(ka)}\frac{dO_\ell(ka)}{dr}\big{|_{r=a}},
\end{equation}
is the dimensionless logarithmic derivative of $O_\ell$ at the channel radius $a$, $L_\ell^*$ is the conjugate of $L_\ell$, and 
\begin{equation}\label{eq26}
	\phi_\ell=\arg I_\ell(ka)=-\arctan\left[\frac{F_\ell(ka)}{G_\ell(ka)}\right],
\end{equation}
is the hard-sphere phase shift.

The internal wave function can then be calculated from $\mathscr{R}_{\ell j}(E)$ and the external wave function as
\begin{equation}\label{eq23}
	\chi_{\ell j}^{\rm int}(k,r)\equiv\psi_{\ell j}^{\rm int}(r)=\dfrac{\hbar^2}{2\mu a \mathscr{R}_{\ell j}(E)} \psi_{\ell j}^{\rm ext}(a)\sum_{j=1}\varphi_j(r)(\mathbf{C}^{-1})_{ij}\varphi_i(a).
\end{equation}

\subsection{The bound state}
The $R$-matrix formalism can be formulated to the bound state problem $(E=E_B < 0)$. The external part of the bound state wave function is the asymptotic form expressed 
explicitly via the Whittaker function \cite{Abra65} as
\begin{equation}\label{eq13}
	\phi_{n\ell j}^{\rm ext}(r)\equiv\psi^{\rm ext}_{n\ell j}(r)=b_{n\ell j}W_{-\eta,\ell+1/2}(2\kappa r), 
\end{equation}
where $b_{n\ell j}$ is the single-particle ANC mentioned in Eq.~\eqref{eq15}, $\kappa$ is the wave number of the bound state.

For the bound state problem, a convenient choice for the boundary parameter $B$ in \eqref{eq18} is \cite{Des10}
\begin{equation}\label{eq30}
	B=S_{\ell}(E_B)=2\kappa a \dfrac{W'_{-\eta,\ell+1/2}(2\kappa r)}{W_{-\eta,\ell+1/2}(2\kappa r)}
\end{equation} 
then, the right-hand side of Eq. \eqref{eq19} is suppressed and the Eq. \eqref{eq20} becomes
\begin{equation}\label{eq31}
	\sum_{n=1}^NC_{in}(E,B)c_n=0.
\end{equation}
This system of equations is similar to a standard eigenvalue problem, but the parameter $B$ depends on the energy $E_B$. In practice, we starts from $B=S_{\ell}(E_B) = 0$ and iteratively solve the system equations \eqref{eq31} until the energy $E_B$ converges. At convergence, the coefficient $c_n$ can be obtained. 

The bound state wave function satisfies the normalization condition
\begin{equation}
	\int_0^\infty |\psi_{n\ell j}(r)|^2 dr =1. \label{eq14}
\end{equation}
To satisfy this condition, one has to add a coefficient $N_{\ell}$ as
\begin{equation}\label{eq32}
	N_{\ell}=1+\gamma^2_{\ell}\left[\dfrac{dS_{\ell}(E_B)}{dE}\right]_{E=E_B},
\end{equation}
where $\gamma_{\ell}$ is the reduced width amplitude of the bound state determined as
\begin{equation}\label{eq33}
	\gamma_{\ell}=\left(\dfrac{\hbar^2}{2\mu a}\right)^{1/2}\sum_{n=1}^Nc_n\varphi_n(a).
\end{equation}
Then the internal wave function of bound state is given by \eqref{eq16} multiplied by $N_{\ell}^{-1/2}$
\begin{equation}\label{eq34}
	\phi_{n\ell j}^{\rm int}(r)\equiv\psi_{n\ell j}^{\rm int}(r)=N_{\ell}^{-1/2}\sum_{n=1}^{N}c_n\varphi_n(r).
\end{equation}
Applying the continuity condition at the boundary $r=a$ and using Eqs.~\eqref{eq13} and \eqref{eq34}, the single-particle ANC determining the external part of the bound state wave function in Eq.~\eqref{eq13} is given by 
\begin{equation}\label{eq35}
	b_{n\ell j}=N_{\ell}^{-1/2}\sum_{n=1}^{N}c_n\varphi_n(a)/W_{-\eta,\ell+1/2}(2\kappa a).
\end{equation}

\section{Results and discussion}
\label{sec4}

\subsection{$^{16}$O(p,$\gamma)^{17}$F reaction} \label{sec41}

\begin{table}[pt]
		\tbl{Potential parameters, single-particle ANCs, and spectroscopic factors used in the $^{16}$O($p,\gamma$)$^{17}$F calculation. \label{t4}}{
			\begin{tabular}{@{}ccccccc@{}}
				\toprule
				$-E_{B}$&Potential & $V_{\rm L(NL)}$ & $R_{\rm c}$&$a_{\rm c}$& 	$b_{n_f\ell_fj_f}$ & $S_{\rm F}$\\
				(MeV)& &  (MeV)&(fm)&(fm)& $(\text{fm}^{-1/2})$ &\\
				\colrule
				\multicolumn{7}{c}{Ground state ($5/2^+$)} \\ 
				0.600&Local & 53.66 & 3.150 &0.65& 0.96 &1.07  \\
				&Nonlocal & 67.86   & 3.150 &0.65 & 1.11 &0.85  \\
				\multicolumn{7}{c}{Excited state ($1/2^+$)} \\
				0.105&Local & 53.02 & 3.150 &0.65& 81.51 & 0.91 \\
				&Nonlocal & 68.28   & 3.150 &0.65 & 88.09 & 0.80 \\
				\multicolumn{7}{c}{Scattering state}  \\ 
				&Local & 53.66 & 3.150 &0.65& & \\
				&Nonlocal & 67.86   & 3.150 &0.65 & & \\
				\botrule
		\end{tabular}} 	 
\end{table}

First, we consider the $^{16}$O($p,\gamma)^{17}$F reactions leading to the ground ($5/2^+$) and excited ($1/2^+$) states in the energy range $E_{\rm c.m.}=0.0-2.5$ MeV. The $S$-factor of these reactions are E1 dominant \cite{Bay98,Des99,Mor97} with the incoming proton in $p$-wave and exhibit a smooth nonresonant behavior under 2.4 MeV. In the considered energy range, the $^{16}$O($p,\gamma)^{17}$F reaction is a well-known example of a nonresonant direct capture process, with the proton captured by a doubly-magic $^{16}$O core forming the bound and first excited states of $^{17}$F nucleus. The experimental data for these transitions are measured with good accuracy, especially in Ref.~\citen{Mor97} (see also Ref.~\citen{Ili08} for a review of existing data). These data show that the transition to the weakly bound ($E_B\approx -0.105$ MeV) $1/2^+$ excited state of $^{17}$F displays an interesting mechanism where it is much stronger than the transition to the ground state \cite{Rol73,Mor97}. This is due to the proton-halo nature of the $1/2^+$ excited state, whose wave function contains a distinctive long tail \cite{Mor97,Bay98}. These features make the $^{16}$O($p,\gamma)^{17}$F reaction a suitable subject to investigate the nonlocality effect in the potential and demonstrate the reliability of the Lagrange-mesh $R$-matrix method.  

The $^{16}$O($p,\gamma)^{17}$F reaction is an extranuclear capture process at low energies, where the reaction takes place at a large distance between the proton and $^{16}$O core. This means that no compound nucleus resonance is involved in the calculation. On the other hand, the integration in Eq.~\eqref{eq3} needs to be done in a large enough radius. Therefore, we choose the channel radius $a=20$ fm and the number of bases $N=150$ to ensure a proper convergence for the results. The calculation with this number of basis $N$ can be done very fast due to the simple single-channel nature considered in our calculation. The $R$-matrix method in this work can be naturally extended to the coupled-channels calculation with nonlocal potential in the future.   

To calculate the radiative capture $S$-factor, we use the WS shape for the local and nonlocal central potentials as described in Section \ref{sec2} with parameters presented in Table \ref{t4}. The parameters of the local spin-orbit part, which is used with both local and nonlocal central potential, are $V_{\rm SO} = 5$ MeV, $R_{\rm SO}=1.25A^{1/3}$, and $a_{\rm SO} = 0.65$ fm. In general, these potentials have standard shapes, which are similar to those used in elastic scattering \cite{Jag18} and radiative capture \cite{Hua10,Tia18}. For the nonlocal potential, the nonlocality range $\beta$ for the proton projectile is 0.85 fm \cite{Per62}. As with all potential model calculations of radiative capture, we consider the $^{17}$F nucleus to be in a simple valence nucleon plus core structure, with the proton in the $5/2^+$ ground state and $1/2^+$ excited state occupying the $1d_{5/2}$ and $2s_{1/2}$ orbitals, respectively. For both the local and nonlocal cases, we use the same potential in the calculations of the scattering- and the ground-states wave functions. The strengths of the potentials for the ground and excited states are varied to produce the experimental values \cite{Til93} of proton separation energy $-E_B$. The spectroscopic factors are fine-tuned independently for the local and nonlocal cases to obtain the best fit in terms of $\chi^2$ with experimental data below 2.5 MeV. We note that the fitted values of the spectroscopic factors shown in Table \ref{t4} are within the range of those reported in the literature \cite{Rol73,Dub17}, although those of the nonlocal potentials have more realistic values (0.85 and 0.80 compared to 1.07 and 0.91) in general.    

\begin{figure}[th] 
		\centerline{\includegraphics[scale=0.45]{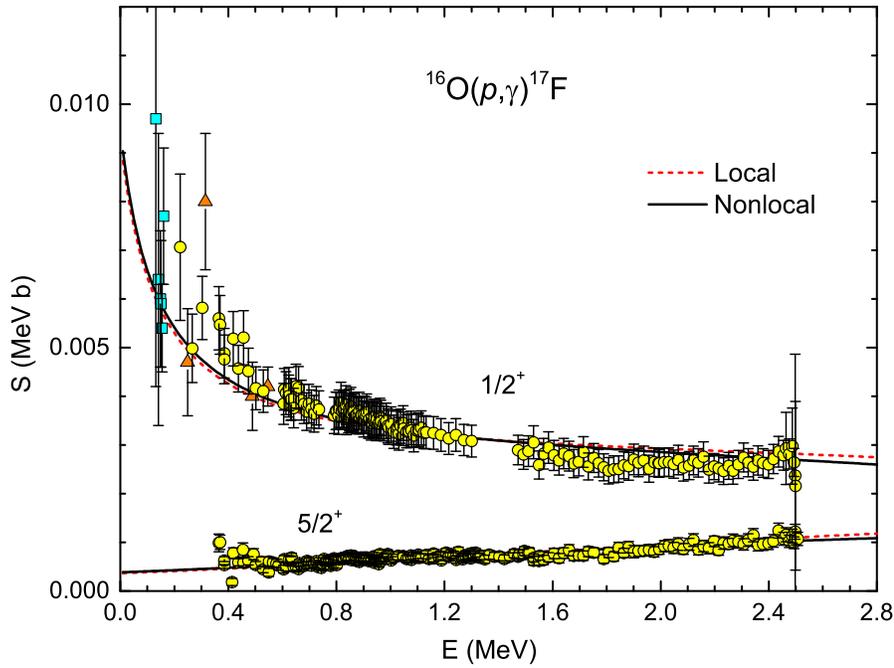}} \vspace*{-0.4cm}
		\caption{Astrophysical $S$-factor of the $^{16}$O($p,\gamma)^{17}$F with local (dashed line) and nonlocal (solid line) potentials. The experimental data shown as squares, triangles, and circles were taken from Refs.~\citen{Hes58}, \citen{Tan59}, and \citen{Mor97}, respectively.} \label{fig5}
\end{figure} 

The theoretical astrophysical $S$-factors of $^{16}$O($p,\gamma)^{17}$F reaction for the ground and excited states of $^{17}$F are compared with the experimental data \cite{Hes58,Tan59,Mor97} in Fig.~\ref{fig5}. The data of Ref.~\citen{Mor97} used in our calculations are the those reevaluated and corrected by Iliadis \emph{et al.} \cite{Ili08}. The calculated results for both local and nonlocal potentials are in good agreement with experimental data for the transition to the ground state and the lower energy region of the excited state. For the $S$-factor of the excited state with center-of-mass energy $E>1.4$ MeV, both results slightly overestimated with the data. This small overestimation is similar to previous potential model calculations \cite{Gag99,Mor97}. 

From the $S$-factor of the excited state in Fig.~\ref{fig5}, we see that the result with the nonlocal potential is in better agreement with data compared to the one with local potential, albeit the largest difference between the calculated results in this energy region is less than $5\%$. This is due to the nonlocality effect mostly manifest in the interior region of the wave function, to which the $S$-factor of nonresonant reactions is not very sensitive. It is important to note that although the difference between the results of local and nonlocal potentials in our work is much smaller than those observed in Ref.~\citen{Tia18}, in contrast to us the authors of that work used the same spectroscopic factor for the calculation of both types of potential. Using the same spectroscopic factors for local and nonlocal potentials in our work results in a similar difference observed in Ref.~\citen{Tia18}. However, since the spectroscopic factor is not a uniquely defined quantity for peripheral reaction \cite{Muk97,Muk01,Muk10}, there is no justified reason to fix it for both types of potential in our calculation. As can be seen later on, allowing the spectroscopic factor to be fitted with experimental data as commonly done in other potential models studies \cite{NACRE2,Hua10} as well as ours leads to different conclusions regarding the impact of nonlocality effect from those in Ref.~\citen{Tia18}.  

\begin{table}[pt]
	\tbl{Data on the astrophysical $S$-factors and ANCs of $^{16}$O($p,\gamma$)$^{17}$F reaction. \label{t5}}{
	\begin{tabular}{@{}llllll@{}} \toprule
		&       & {$S(0)_{d5/2}$} & {$S(0)_{s1/2}$ } & {$C_{d5/2}$ } & {$C_{s1/2}$ } \\ 
		&&(keV b)&(keV b)&($\text{fm}^{-1/2}$)&($\text{fm}^{-1/2}$) \\ \colrule
		Present work (local) &       & 0.36  & 9.12   & 0.99  & 77.76 \\
		Present work (nonlocal) &       & 0.39  & 9.32   & 1.02  & 78.79 \\
		Garliardi \emph{et al.} 1999 \cite{Gag99} &       & {$0.40 \pm 0.04$} & {$9.8 \pm 1.0$} & {$1.04\pm0.05$} & {$80.56\pm4.22$} \\
		Baye \emph{et al.} 1998 \cite{Bay98} &       & 0.45  & {9.76--10.52} & 1.1   & \multicolumn{1}{l}{78.6--81.6} \\
		Bing \emph{et al.} 2007 \cite{Bin07} &       & {$0.35\pm0.05$} & {$9.3 \pm 1.4$} & {$0.99\pm0.07$} & {$81.97\pm6.03$} \\
		Artemov \emph{et al.} 2009 \cite{Art09} &       & {$0.40 \pm 0.04$} & {$9.37 \pm 0.36$} & {$1.04\pm0.05$} & {$75.50\pm1.49$} \\
		Huang \emph{et al.} 2010 \cite{Hua10} &       & 0.304 & 9.075 & 0.91  & 77.21 \\ \botrule
	\end{tabular}}%
\end{table}%

Other essential quantities to evaluate the two types of potential are the $S$-factor value at zero-energy limit $S(0)$ and the ANC $C_F$ value. These quantities for the ground and excited states transitions calculated in the present work are compared with those reported in the literature in Table \ref{t5}. The ANCs are particularly important for nonresonant capture reactions where most contribution coming from the asymptotic tail region of the integrand in Eq.~\eqref{eq3}. The $S(0)$ and $C_F$ values of both the local and nonlocal potential cases are in good agreement with those reported in previous works. It is interesting to note that although the single-particle ANCs $b_{nlj}$ and the spectroscopic factors of the local and nonlocal bound state are different by up to 20\%, these differences cancel out in the ANC $C_F$, which is a more appropriate quantity to characterize the nonresonant capture process \cite{Muk97,Muk01}.  

The total $S(0)$ values, which are 9.48 keV b and 9.71 keV b for the local and nonlocal calculations, respectively, also agree with those reported in important reaction rates evaluations such as $9.3 \pm 2.8$ keV b \cite{Ang99} and $10.6 \pm 0.8$ keV b \cite{Ade11}. The difference between the $S(0)$ values of the local and nonlocal calculations is just 2\%. Our calculations for the $^{16}$O($p,\gamma)^{17}$F reaction with the local and nonlocal potentials suggested that for the nonresonant direct capture reactions in general, the use of the nonlocal potential can slightly improve the result, although the difference between the two results is much smaller than the usually assumed theoretical uncertainty of 10\%.

\subsection{$^{13}$C(p,$\gamma)^{14}$N reaction} \label{sec42}
In this part, we applied the same Lagrange-mesh $R$-matrix method to the $^{13}$C($p,\gamma_0)^{14}$N reaction leading to the $1^+$ ground state of $^{14}$N, which has a distinctive feature compared to the nonresonant $^{16}$O($p,\gamma)^{17}$F reaction in the previous calculation. The experimental $S$-factor data of the $^{13}$C($p,\gamma_0)^{14}$N reaction in the center-of-mass energy range $E=0-1.8$ MeV display two remarkable E1 resonances \cite{Heb60,Kin94,Zep95,Gen10}. They are the relatively broad $1^-$ ($E_R=0.518$ MeV, $\Gamma=0.037$ MeV) and $0^-$ ($E_R=1.225$ MeV, $\Gamma=0.408$ MeV) resonances \cite{Kin94}, which we denote R1 and R2, respectively. A simultaneous description of both of these resonances is particularly challenging that it has only been done in phenomenological $R$-matrix \cite{Cha15} or hybrid Breit-Wigner approaches \cite{Cha19,Zep95,Li12,Kab20}. Such simultaneous description for the $^{13}$C($p,\gamma_0)^{14}$N $S$-factor with a pure potential model has not been reported before. 

\begin{table}[pt]
	\tbl{Potential parameters and spectroscopic factors used in the $^{13}$C($p,\gamma_0)^{14}$N reaction. \label{t6}}{
		\begin{tabular}{@{}cccccc@{}}
			\toprule 
			$-E_B$($E_R)$&Potential & $V_{\rm L(NL)}$ & $R_{\rm c}$&$a_{\rm c}$& $S_{\rm F}$\\
			(MeV)& &  (MeV)&(fm)&(fm)&\\
			\colrule
			\multicolumn{6}{c}{Ground state ($1^+$)} \\ 
			7.550&Local & 52.26 & 2.939 &0.65&  \\
			&Nonlocal & 61.89   & 2.939 &0.65 & \\
			&&&&&\\
			\multicolumn{6}{c}{Resonance state}  \\ 
			\multicolumn{6}{c}{Standard form}  \\ 
			0.51&Local & 53.85 & 2.939 &0.65& 0.22 \\
			&Nonlocal & 69.50   & 2.939 &0.65& 0.19 \\
			1.22&Local & 47.83 & 2.939 &0.65& 0.55\\
			&Nonlocal & 60.00   & 2.939 &0.65 & 0.44\\
			&&&&&\\
			\multicolumn{6}{c}{Narrow form}  \\ 
			0.51&Local & 24.42 & 1.944 &0.04& 0.21 \\
			&Nonlocal & 27.87   & 1.944 &0.04 & 0.19\\
			1.22&Local & 22.10 & 1.944 &0.04& 0.35 \\
			&Nonlocal & 25.00   & 1.944 &0.04 & 0.31\\
			\botrule
	\end{tabular}}  
\end{table}   

\begin{figure}[th] 
	\centerline{\includegraphics[scale=0.45]{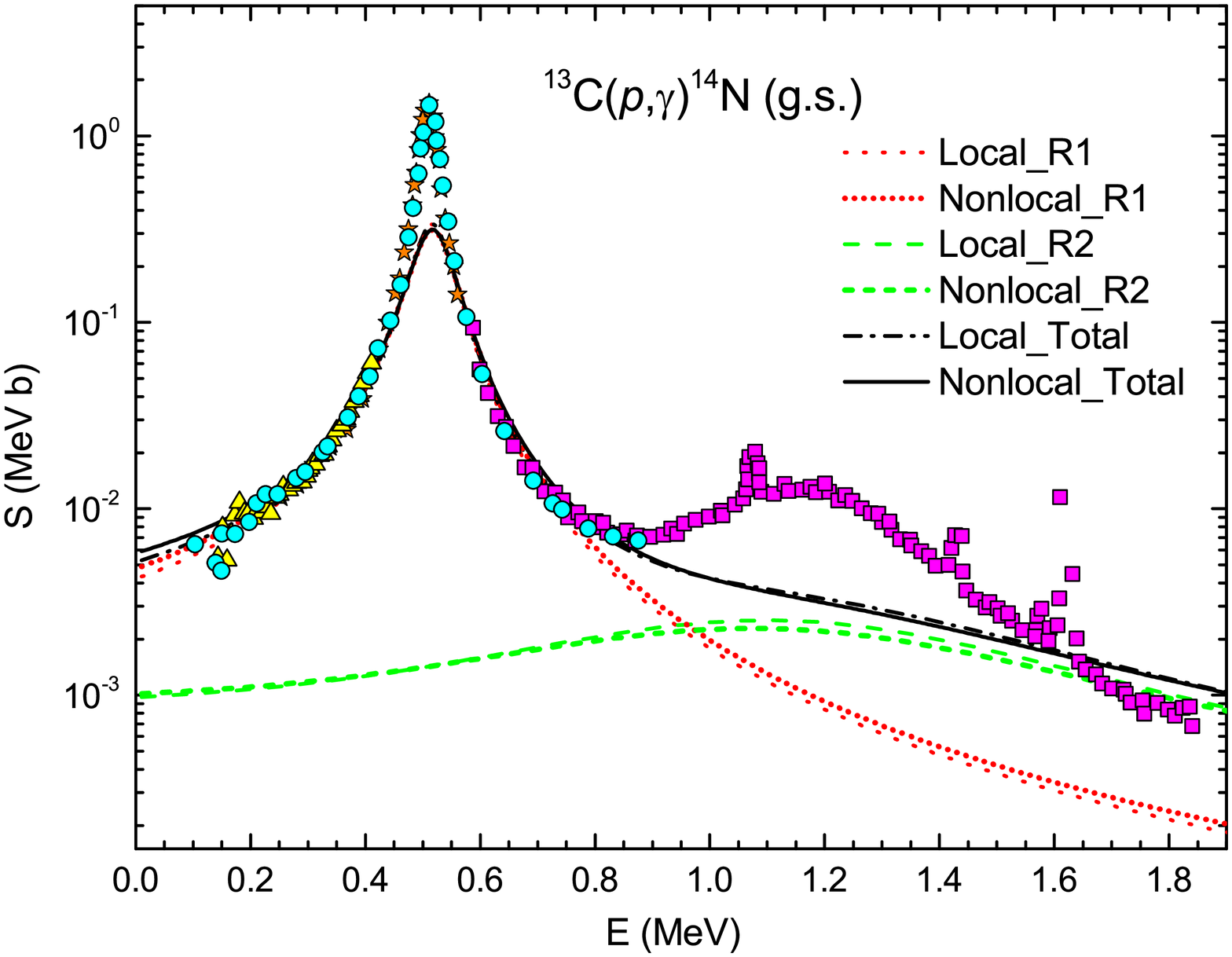}} \vspace*{-0.4cm}
	\caption{Astrophysical $S$-factor of the $^{13}$C($p,\gamma_0)^{14}$N reaction with local (dashed-dotted line) and nonlocal (solid line) potentials of ``standard'' form. The contributions from the R1 and R2 resonances with local (dotted and dashed lines) and nonlocal potentials (short-dotted and short-dashed lines) are also shown. The experimental data shown as stars, circles, triangles, and squares were taken from Refs.~\citen{Gen10}, \citen{Kin94}, \citen{Heb60}, and \citen{Zep95}, respectively.} \label{fig2}
\end{figure}       

In this calculation, the ground state of $^{14}$N is formed by coupling the bound $1p_{1/2}$ valence proton to the $1/2^-$ $^{13}$C core with $E_B=-7.55$ MeV. Due to the weak contributions from higher $\ell_i$ components in the reactions of interest \cite{Kin94,Muk03}, we only consider the scattering wave function with $\ell_i=0$. The transition from this scattering $s$-wave to the ground state is dominant by the E1 component. In the considered energy regime besides the two mentioned broad resonances, the $S$-factor data also exhibit several very narrow resonant peaks \cite{Cha19}. However, these peaks do not have a considerable contribution in the $S$-factor at astrophysical-interesting energy. Thus, for the purpose of the present study, we only consider the transition from the R1 ($1^-$, 0.51 MeV) and R2 ($0^-$, 1.22 MeV) resonances. For the specific $^{13}$C($p,\gamma_0)^{14}$N reaction with $\ell_i=0$ scattering wave, these  $1^-$ and $0^-$ resonances are results of the couplings between the spin $1/2$ incoming proton and the $1/2^-$ $^{13}$C target. We note that for a general radiative capture reaction, there exists also a nonresonant contribution, which is generated using the same potential as the bound state and constrained by the ANC in the same manner as the procedure discussed in Sec.~\ref{sec41}. Using the same criteria as NACRE II for selecting nonresonant and resonant contributions \cite{NACRE2}, we do not include such nonresonant component as the combination of $J_i$ and $\ell_i$ have already been used up by the two resonant transitions. The spectroscopic factors reported in Table \ref{t6} are also used as the fitting parameters in order to obtain the correct resonance height. It has been discussed in Refs.~\citen{Anh21,NACRE2} that the spectroscopic factors used to describe resonant capture processes in the potential model do not represent a physical quantity and cannot be compared with those obtained from microscopic calculations or transfer reactions analyses.    

A distinctive difference of the $^{13}$C($p,\gamma)^{14}$N calculation compared to the $^{16}$O($p,\gamma)^{17}$F one is the use of multiple potentials to generate the scattering wave functions corresponding to each of the resonances. The depths of these potentials are separately adjusted to reproduce the correct position and width of the resonance. This type of resonant parametrization in radiative capture reactions has been commonly used in most potential model studies \cite{NACRE2,Dub12}. As mentioned earlier, the  $1^-$ and $0^-$ resonances are results of the couplings between the spin $1/2$ incoming proton and the $1/2^-$ $^{13}$C target. Thus, the use of two separated potentials can be interpreted as an approximate way to account for the effect of the spin-spin interaction between the proton and $^{13}$C \cite{Prz90}. A quantitative investigation of the effect of an explicit spin-spin term in the nuclear potential (as in Ref.~\citen{Prz90}) on the direct capture cross section is of interest but beyond the scope of this work.   

We perform the single-channel Lagrange-mesh $R$-matrix with the same number of basis $N=150$ and the channel radius $a=20$ fm as the previous case. First, we consider a ``standard'' form for the shape of the WS potentials for both local and nonlocal cases. Table \ref{t6} shows the standard potential parameters, whose shapes are similar to the those in previous Sec.~\ref{sec42} and other works \cite{Hua10,Anh21,Kab20,Cha19}. The spin-orbit potential and the range of nonlocality are the same as in Sec.~\ref{sec41}. The calculated $S$-factor is compared with the experimental data in Fig.~\ref{fig2}. Here we omitted the error bars since their size is comparable with the data point. Fig.~\ref{fig2} shows clearly that the results with local and nonlocal potentials in the standard form cannot describe the R2 resonance. Both calculations also cannot fill the experimental R1 resonant peak, which has been pointed out in previous potential model analyses using the standard shape potential \cite{Hua10,Anh21}. Careful checks have been done by varying the potential depth from 0 to 100 MeV without any success in describing these resonances with the standard shape. 

\begin{figure}[th] 
	\centerline{\includegraphics[scale=0.45]{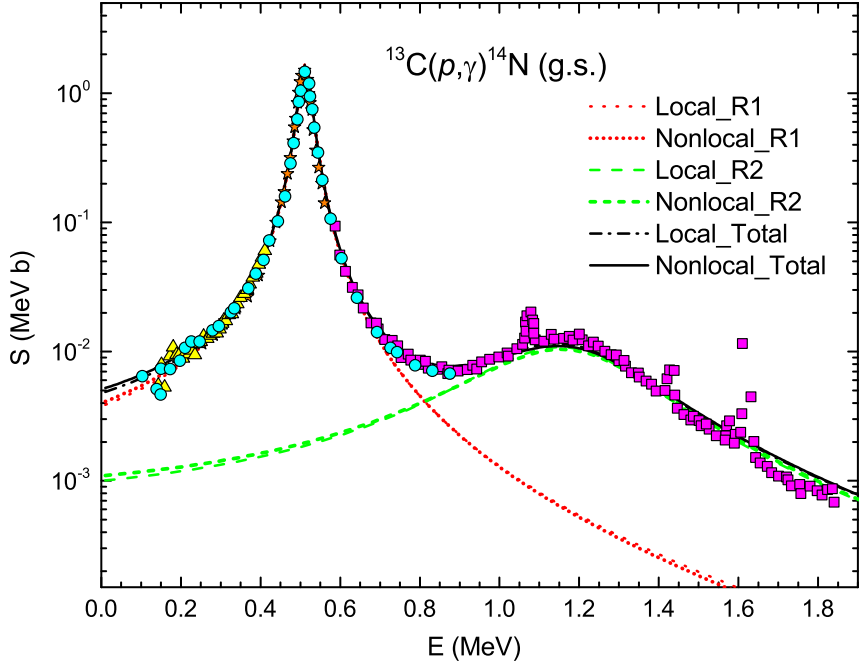}} \vspace*{-0.4cm}
	\caption{ Similar to Fig.~\ref{fig2} but with the ``narrow'' form for the local and nonlocal potentials used to calculate the scattering wave functions.} \label{fig3}
\end{figure} 

As has been suggested in other potential model studies \cite{NACRE2,Dub12}, the peak of the R1 resonance can be satisfactorily described by using an unconventional narrow (i.e. small radius) and shallow shape for the scattering potentials. These potentials can have the almost square well shape with very small diffuseness \cite{NACRE2} or Gaussian form \cite{Dub12} (see also Fig. 7 of Ref.~\citen{Anh21}). Therefore, in this study, we adopt the potential shape in Ref.~\citen{NACRE2} for both the local and nonlocal potentials used to generate the scattering wave functions. The potential parameters are displayed in the lower part of Table \ref{t6}. The calculated $S$-factor with these so-called ``narrow'' potentials is shown in Fig.~\ref{fig3}. We obtain a very good agreement between the theoretical results and the experimental data for both the local and nonlocal cases. To our knowledge, this is the first reported simultaneous description of both R1 and R2 resonances in $^{13}$C($p,\gamma_0)^{14}$N using only the potential model. 

The use of a somewhat unconventional narrow shape for the description of multiple resonances in the cross section can be justified as follows. In the potential model description of radiative capture reactions, the resonance in the cross section or $S$-factor is obtained with a scattering wave function with distinctive features different from the bound and nonresonant wave functions \cite{She85,Muk10-2}. The narrow potentials in this study act solely as the resonance generator similar to manually adding resonance pole in the phenomenological $R$-matrix method, thus these potentials are not required to possess a standard shape that is used to describe the bound state and nonresonant wave functions. We remark that our use of multiple narrow potentials for each resonance shares the same idea as the modified potential cluster model of Dubovichenko \emph{et al.} \cite{Dub12}.  

\begin{table}[pt]
	\tbl{Data on the astrophysical $S$-factors at zero energy of $^{13}$C($p,\gamma_0)^{14}$N reaction. \label{t7}}{
	\begin{tabular}{@{}ll@{}} \toprule
		& $S(0)$ \\
		& keV b \\ \colrule
		Present work (local) & 4.78 \\
		Present work (nonlocal) & 5.12 \\
		King \emph{et al.} 1994 \cite{Kin94} & 5.25 \\
		Mukhamedzhanov \emph{et al.} 2003 \cite{Muk03} & $5.16 \pm 0.72$ \\
		Artemov \emph{et al.} 2008 \cite{Art08} & 5.06 \\
		Genard \emph{et al.} 2010 \cite{Gen10} & $3.94 \pm 0.59$ \\
		Huang \emph{et al.} 2010 \cite{Hua10} & 6.22 \\
		Li \emph{et al.} 2012  & $5.78 \pm 0.48$ \\
		Chakraborty \emph{et al.} 2015 \cite{Cha15} & $4.72 \pm 0.86$ \\
		Chakraborty \emph{et al.} 2019 \cite{Cha19} & $4.97 \pm 0.77$ \\
		\botrule
	\end{tabular}}
\end{table}%

From the results in Figs.~\ref{fig2} and \ref{fig3}, one sees that the difference in the calculated results between the local and nonlocal potentials is very small (less than 8\%). This situation is similar to the $^{16}$O($p,\gamma)^{17}$F case in Sec.~\ref{sec41}. Table \ref{t7} compares the calculated $S$-factor at zero energy limit with those of previous works. The $S(0)$ for both local and nonlocal cases are completely within the uncertainty range of the reported values in the literature. We conclude that for the $^{13}$C($p,\gamma_0)^{14}$N case with two broad resonances the local and nonlocal potentials provide appropriate results with discrepancy well under the usually adopted uncertainty. 

From the $^{13}$C($p,\gamma_0)^{14}$N and $^{16}$O($p,\gamma)^{17}$F cases we can explain the small difference in the results between the local and nonlocal potentials as follows. The general effect of a Perey-Buck-type nonlocal potential is to modify the interior part of the wave function \cite{Per62}, which leads to a difference in the same part of the integrand in Eq.~\eqref{eq3}. However, with the normalization of the calculated cross section with the experimental data by adjusting the spectroscopic factor, which is a common practice of most potential model analyses, the exterior part of the integrand \eqref{eq3} is compensated by an averagely comparable amount to the difference observed in the interior part. This interplay between the differences within each side of the integrand reduces the nonlocality effect observed on the normalized radiative capture cross section or $S$-factor. We remark that our finding does not conflict with the results in Ref.~\citen{Tia18}, it only means that with the usual practice of normalizing the cross section to experimental data done in most potential studies, the effect of the potential nonlocality is likely to be small.

\section{Summary}
\label{sec5}
In this study, we applied the Lagrange-mesh $R$-matrix method to investigate the impact of nonlocality effect in the nuclear potential on the nonresonant $^{16}$O($p,\gamma)^{17}$F and resonant $^{13}$C($p,\gamma$)$^{14}$N capture reactions. Using the narrow scattering potentials for the $^{13}$C($p,\gamma_0$)$^{14}$N reaction, we also obtain for the first time a simultaneous description of the two resonances for $E < 1.8$ MeV using a pure potential model. 

The theoretical $S$-factors and ANCs are compared with experimental data and reported value from previous works with very good agreement. The use of nonlocal potential improves the results but in contrast to the finding of Ref.~\citen{Tia18}, we found that the difference in the results between the local and nonlocal potential is at most 8\%, which is smaller than the uncertainties originated from the experimental procedure or other theoretical assumption for most capture reactions. This is due to the normalization procedure usually done in potential model studies. Nevertheless, we expect that the nonlocality effect could play a more important role for key capture reactions with high accuracy experimental data such as $^{7}$Be($p,\gamma)^{8}$B and $^{12}$C($\alpha,\gamma)^{16}$O. We also propose that future nuclear astrophysics reaction rates evaluations consider the uncertainty coming from the nonlocality effect in their works.

Finally, we note that the nonlocal potential in this study assumes a phenomenological Perey-Buck form. More realistic nonlocal microscopic nucleon-nucleus potentials can be readily obtained with the folding model \cite{Amo00,Hao15,Rot17,Loa20}. These potentials can provide more consistent and reliable results for the radiative capture reactions. Moreover, the Lagrange-mesh $R$-matrix framework used in this study can be extended to incorporate these nonlocal potentials in a coupled-channel description of capture reaction. Studies of radiative capture reactions in these research lines are currently underway.

\section*{Acknowledgements}
We thank Pierre Descouvemont for his communication on the numerical calculation 
of the \pG cross section and for providing the data for $^{16}$O($p,\gamma)^{17}$F reaction. We also thank Dao Tien Khoa for his enlighten discussions. The present research was supported by Vietnam Atomic Energy Institute (VINATOM) under the grant 
{\DJ}TCB.01/19/VKHKTHN. 

\bibliographystyle{bibformat}
\bibliography{sample}
\end{document}